\documentstyle{l-aa}
\input epsf

\begin{document}

\thesaurus{08         
              (01.01.2;
		02.02.1;  
               13.07.2;  
               08.02.3; 
	       13.25.3)}
\title{RXTE broad band observations of X-ray Nova XTE J1755-324} 

\author{M. Revnivtsev\inst{1,3}, M. Gilfanov\inst{2,1} and E.
Churazov\inst{2,1}}

\offprints{M. Revnivtsev}

\institute{ Space Research Institute,Russian Academy of Sciences
Profsoyuznaya 84/32, 117810 Moscow, Russia
\and Max-Planck-Institut f\"ur Astrophysik,
Karl-Schwarzschild-Str. 1, 85740 Garching bei Munchen, Germany
\and visiting Max-Plank-Insitute f\"ur Astrophysik}

\date{}

\maketitle

\begin{abstract}
The properties of X-ray Nova XTE J1755--324, observed with RXTE 
observatory in 1997 are reported. The lightcurve of the 
source was typical for X-ray Novae, except for somewhat shorter decay
time scale and the time elapsed between the primary and subsequent
secondary and tertiary outbursts. At 
the peak of the lightcurve the source had two-component spectrum, typical for
the bright X-ray Novae, with a characteristic disk blackbody temperature
$T_{in}\sim0.8$ keV and a photon index of the hard power law tail $\alpha\sim$
2.0. The peak luminosity can be roughly estimated as $L\sim10^{38}$ erg/s
(0.1--100 keV, assuming 8.5 kpc distance). 
A notable and peculiar feature of the spectral evolution of the source
was a short, $\sim 10$ days long, episode of increased hardness,
occurred shortly before the tertiary maximum of the light curve.
During the last RXTE pointed observations $\sim 100$ days after the primary
outburst the source  has been found in the hard spectral state with
luminosity $\sim$few$\times 10^{36}$ erg/s. The overall pattern of the temporal
and spectral evolution of XTE J1755--324 resembles in general that of
``canonical'' X--ray Novae (e.g. Nova Muscae 1991) and suggests that
the compact object in the binary system is a black hole.

   \end{abstract}

\markboth{M.~Revnivtsev et. al.: RXTE broad band observations of X-ray Nova XTE
J1755-324}{}
%

\section{Introduction}

XTE J1755--324 was discovered by All--Sky Monitor (ASM) aboard Rossi X-ray
Timing Explorer (XTE) on July 25, 1997 and localized by  
Proportional Counter Array (PCA) at the position with coordinates
R.A.=17$^h$55$^m$28$^s$.6, Dec.=--32\fdg28\arcmin39\arcsec (J2000). 
The light curve of the source was typical 
for X-ray Novae (e.g. X-ray Nova Muscae 91, X-ray Nova Vulpeculae 88, see
\cite{tanaka} for a review) with a quick rise within a few days and an
exponential decay with an e--folding time of $\sim$30--40 days. The spectrum of the
source at the peak of the lightcurve was complex, with a soft component
having approximately multicolor blackbody disk spectrum with
$T_{in}\sim0.8$ keV and a hard power law tail extending above 10 keV. 
Preliminary analysis of the data of BeppoSAX observations of the Galactic
Center region in Sep. 1997 has not revealed X-ray bursts from the source 
(Ubertini 1997, private communication). 
Radio observations of the region of the sky containing the source performed
on Aug 18, 1997 have not found significant radio emission from
XTE J1755--324 down to $\sim0.3$ mJy (\cite{radio}). The optical extinction at
the direction of the source corresponding to $N_H\sim3.7\times10^{21}$
cm$^{-2}$ (\cite{nh}) is $A_v\sim$2, however to our knowledge no
observations of the source at optical wavelengths have been reported.
 
\section{Instruments and observations}

For the analysis presented below we have used the public domain data from 
all three instruments aboard Rossi X-ray Timing Explorer (\cite{rxte}) --
Proportional 
Counter Array (PCA), High Energy X-ray Timing Experiment (HEXTE) and All Sky
Monitor (ASM). The two pointed instruments PCA and HEXTE
provided the broad band spectral data covering an energy range from
3 to $\sim 200$ keV, while the ASM data give a possibility to follow
the long term behavior of the  source with nearly complete time coverage.

XTE J1755--324 was observed by RXTE pointed instruments on several
occasions during 
Summer--Fall 1997. The observations used in the analysis are listed in Table
\ref{obslog}.  The HEXTE data for the 3$^{rd}$ observation were
excluded from the analysis because of a short live time of the observation.

The data were retrieved from the XTE GOF at GSFC. The ASM light curves
were used as provided by the XTE GOF.
The PCA and HEXTE data were analyzed using the standard FTOOLS
(version 4.1) tasks.

The latest version of the PCA background estimator (v.1.5) with the
VLE (Very Large Event) based background model was used (Stark 1997).
For the spectral analysis of the PCA data the version 3.3 of the PCA
response matrix was used (Jahoda 1998). To roughly account for
the uncertainty in the knowledge of the spectral response, a  1\%
systematic error was added to the statistical error in each PCA channel.
The spectral data below 3 keV and above 30 keV were excluded from the
analysis due to increasing systematic uncertainty in these energy
ranges. 
The PCA dead time fraction was calculated following Zhang \& Jahoda
(1996). Typical values of the
dead time fractions were $\sim 2.5-3.0\%$.

\begin{table}
\small
\caption{The list of RXTE observations of XTE J1755--324. \label{obslog}}
\begin{tabular}{cccccc}
\hline
&Obs.ID&&&\multicolumn{2}{c}{Live time$^a$, s}\\
&20425-01-...& Date&UT start&PCA&HEXTE$^b$\\
\hline
1&01& 29/07/97  & 05:52:48   &  767 & 360\\
2&02-00    & 01/08/97  & 05:19:12  & 1993 &  652\\
3&03-00A   & 06/11/97  & 15:00:00  &   16 & -\\
4&03-03S   & 06/11/97  & 15:34:02  &  652 & 735\\
5&04-00    & 12/11/97  & 15:38:08  & 3039 & 1673\\
\hline
\end{tabular}
\begin{list}{}{}
\item[$^a$]-- Deadtime corrected value
\item[$^b$]-- Time for each cluster of detectors
\end{list}
 
\end{table}
\begin{center}

\begin{table*}
\small
\caption{Parameters of spectral approximations of RXTE observations of XTE
J1755--324. \label{param}} 
\begin{tabular}{ccccccccc}
\hline
\\
\multicolumn{8}{c}{Multicolor disk + power law + gaussian line (PCA data)}\\
\\
\hline
\# Obs.&$T_{in}$&$R_{in}^{a}cos(\theta)^{-1/2}$,&$E_{line}^b$&$FWHM_{line}$&$EW_{line}$&$F_{line}$&$\alpha$&$F_{3-30 keV}^c$\\
&keV& km&keV&keV&eV&$10^{-4}$phot/cm$^2$/s&&\\
\hline
1& $0.75\pm0.01$&$35.5\pm0.7$&6.4&$2.92\pm0.47$&$220\pm62$&$14.2\pm6.5$&$2.30\pm0.34$&$1.72\pm0.08$\\
2& $0.74\pm0.01$&$35.7\pm0.7$&6.4&$3.49\pm0.23$&$418\pm73$&$23.3\pm4.0$&$1.83\pm0.10$&$1.59\pm0.07$\\
\hline
\\
\multicolumn{8}{c}{Power law + gaussian line}\\
\\
\hline
&\multicolumn{6}{c}{PCA}&\multicolumn{2}{c}{HEXTE}\\
\hline

&$\alpha$&$F_{3-30 keV}^c$&$E_{line}$&$FWHM_{line}$&$EW_{line}$&$F_{line}$&$\alpha$&$F_{15-100 keV}^c$\\
\hline
3&$2.01\pm0.06$&$0.49\pm0.01$\\
4&$1.96\pm0.01$&$0.48\pm0.01$&$7.00\pm0.23$&$1.38\pm0.73$&$126\pm72$&$3.3^{+1.80}_{-0.96}$&$1.76\pm0.15$&$0.29\pm0.04$\\
5&$1.87\pm0.01$&$0.40\pm0.01$&$6.77\pm0.16$&$1.38\pm0.39$&$121\pm36$&$2.7\pm0.9$&$1.66\pm0.16$&$0.28\pm0.02$\\
\hline
\end{tabular}
$T_{in}$, $R_{in}$ -- the temperature and the radius of the inner
edge of the accretion disk in the multicolor blackbody disk model,
$\theta$ -- the inclinantion angle of the disk plane. $E_{line},
FWHM_{line}, EW_{line}$ and $F_{line}$ -- the center, the width,
equivalent width and the flux of the gaussian line. $\alpha$ - the
power law photon index; the hydrogen column density was fixed at the
Galactic value $NH=3.7\cdot 10^{21}$ cm$^{-2}$. 
\begin{list}{}{}
\item[$^a$]-- assuming 8.5 kpc distance
\item[$^b$]-- The line energy was fixed (see text).
\item[$^c$]-- in the units of $10^{-9}$ erg/s/cm$^2$.
\end{list}
 
\end{table*}

\end{center}

For the spectral analysis of the HEXTE data the version 2.6 (released Mar.
20, 1997) of the response matrix was used. The background for each
cluster of HEXTE detectors was estimated using the off-source observations. 
Only the data above 15 keV were used because of the uncertainties of
the response matrix below this energy. At the high energy end the
spectrum was cut at $\sim 80-150$ keV depending on the brightness of
the source in order to avoid possible influence  due to
systematic uncertainties of the background subtraction. 
The deadtime correction was performed using hxtdead FTOOLS task for all observations.

\section{Temporal and spectral behavior of the source.}
\begin{figure}
\epsfxsize=9.0 cm
\epsffile{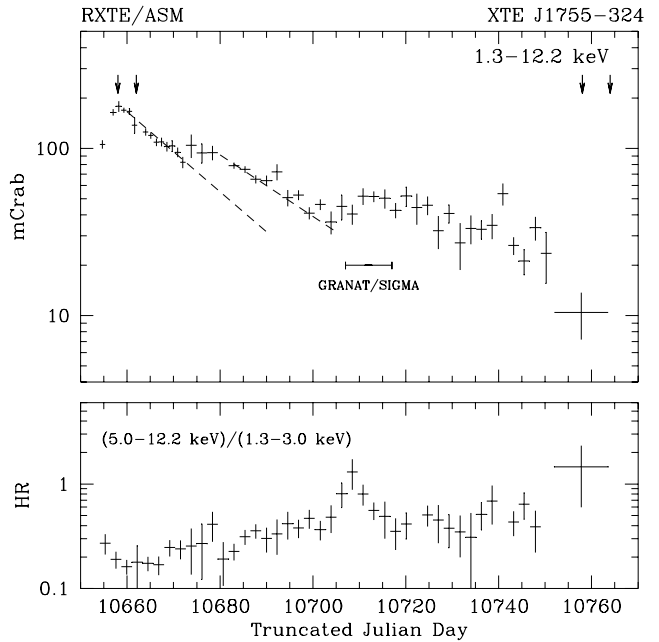} 
\caption{The upper panel: The RXTE/ASM light curve of XTE J1755--324 in
the 1.3--12.2 keV energy band. Vertical arrows mark the dates of RXTE
observations, the solid horizontal line shows the dates of the GRANAT/SIGMA
observations. Dashed lines show 
approximate fit to the light curve before and after the secondary
maximum. The lower panel: The hardness ratio (5.0--12.2 keV)/(1.3--3.0 keV)
derived from the XTE/ASM data. Hardness ratio $\sim$1.0 roughly corresponds
to the Crab-like spectrum.
\label{asmcurve}}   
\end{figure}

\begin{figure*}
\epsfxsize=17 cm
\epsffile[40 220 560 600]{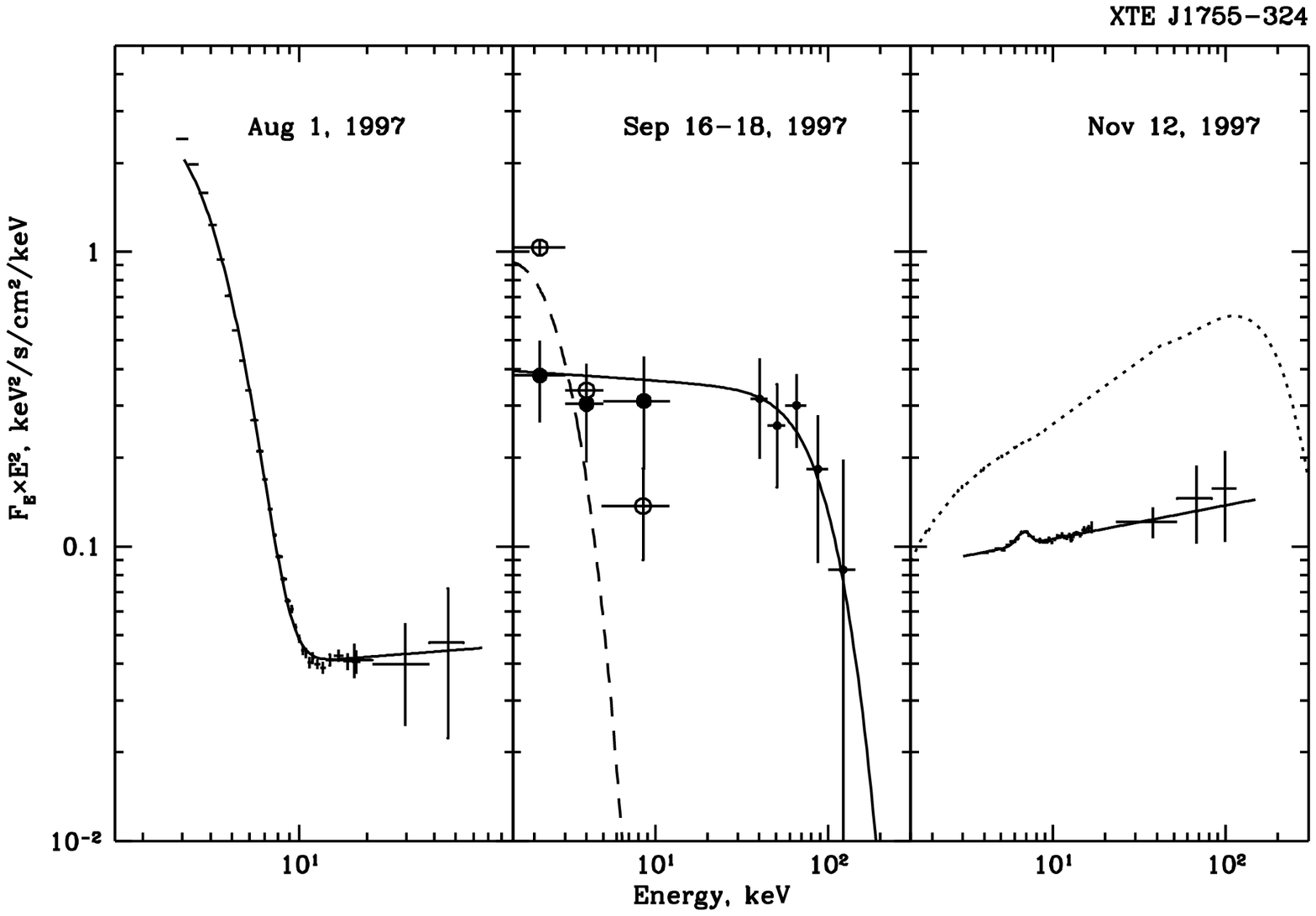} 
\caption{Spectra of XTE J1755--324 observed at different epochs. Data
in the left and right panels were obtained by PCA and HEXTE aboard
RXTE, data in the middle panel were obtained by RXTE/ASM (1.3--12.2 keV)
and GRANAT/SIGMA (35--150 keV, Goldoni et al. 1998, Revnivtsev et al. 1998).
Open circles on the 
middle panel show the approximate spectrum of XTE J1755--324 just before and
after the peak of the hardness (see text). Dashed line on the middle
panel shows the spectrum of the soft component with $R_{in}$ being the
same as during the Aug.1, 1997 observation and $T_{in}$, scaled
according to the luminosity change of the source. Dotted line on the right  panel
is the scaled spectrum of Cyg X-1 in the low spectral state. A solid
line in each panel shows corresponding best fit model spectrum.
\label{spectrum}}    
\end{figure*}

According to the ASM/RXTE data, the light curve of the source was typical for 
X-ray Novae -- a quick (within a few days) rise of the flux was
followed by a much slower decay (Fig.\ref{asmcurve}, the upper
panel). Furthermore,  closer examination of the shape of the light
curve suggests  the presence of two ``kicks'' around TJD 10675
and 10710 (TJD -- Truncated Julian Day = JD--2440000.5) which
resemble the secondary and tertiary maxima observed in the
light curves of several X--ray Novae (see \cite{tanaka} for a review). If
these ``kicks'' are indeed the secondary and tertiary maxima, the XTE
J1755--324 has the shortest time elapsed between the primary and the
subsequent outbursts observed so far -- $\sim$ 15--20 days. The
e-folding decay times between the primary and the secondary maxima and
after the secondary maximum are also among the shortest measured for
the X-ray Novae and equal to $\sim18$ and $\sim28$ days respectively.

The spectrum of the source at the peak of the light curve apparently
consists of two components (Fig.\ref{spectrum}, the left panel). 
Both July 29, 1997 and Aug 1, 1997
pointed observations show that in the standard X-ray band (3--10 keV)
the emission was dominated by the soft spectral component. This soft
component can be approximately fitted by a multicolor blackbody disk model 
(Shakura\&Sunyaev 1973, Mitsuda et al. 1984) with the inner temperature
$T_{in}\sim0.77$ keV and the inner radius of the optically thick part of
the accretion disk $R_{in}\sim30\times(cos\theta)^{-1/2}$ km (assuming
8.5 kpc distance, $\theta$ is the inclination angle of the disk plane).  
At energies higher than $\sim$10 keV the power law component dominates
extending to at least 30 keV with a photon index of $\alpha\sim$ 2.0.

However, the $\chi^2$ value calculated for  Aug.1, 1997 data using two
component  multicolor  blackbody disk + power law model is
rather high $\sim$ 121 (54 dof). Most prominent deviations of the data
from the model are in the $\sim 5-8$ keV energy range and exceed
sufficiently the present calibration uncertainties of the PCA.
Addition of a broad emission line at E=6.4 keV as the third component
of the model improves the quality of the fit in terms of $\chi^2$
statistics ($\chi^2\sim$33 for 52 dof). The change of the central
energy of the line 
to 6.7 keV does not affect the other parameters significantly.
The line flux, 1-2$\times10^{-3}$ phot/s/cm$^2$, sufficiently
exceeds that of the diffuse 6.7 keV line from the Galactic Ridge
approximately derived from the results of GINGA observations of the 
Gacactis Ridge (Yamauchi \& Koyama 1993). 
However, one should bear in mind that due to complex shape of
the spectrum in this energy range and sufficiently large width of the
line, the line parameters depend strongly on the assumed spectral
model for the continuum emission. Moreover, the line itself could be
an artifact caused by  inadequate continuum model. 
Nevertheless, we present in the Table \ref{param}  the best fit
parameters for the model including the emission line.

No statistically significant flux was
detected by HEXTE in the 1$^{st}$ observation with an upper limit 
$F_{15-80~keV}\la1.3\times10^{-10}$ erg/s/cm$^2$, (2$\sigma$). 
The 2$^{nd}$ observation (longer live time) has revealed marginally
statistically significant flux at the level of $\sim1\times10^{-10}$
erg/s/cm$^2$ (15--80 keV, the statistical significance of $\sim 4
\sigma$). The shape of the spectrum obtained by HEXTE is
consistent with the extrapolation of the PCA data -- the HEXTE data 
gives a power law photon index of $\alpha=1.6\pm0.7$.  
It is worth mentioning that the source count rate detected by HEXTE
($\approx 2$ cnt/s) was at the level of $\approx 1.5\%$
of the background count rate ($\approx 120$ cnt/s).
Nevertheless, according to \cite{hexte}, the systematic uncertainty of
the background subtraction using the standard ftools tasks is less than
$\approx 0.5\%$. We may therefore assume that the HEXTE spectrum
shown on the left panel in Fig.\ref{spectrum} is not severely
contaminated by the background.

The spectrum of the source at that time was similar to the spectra of
some X-ray Novae around the maximum of the light curve, e.g. X-ray Nova Vulpeculae 88 
(\cite{mir-kvant}), Nova Muscae 1991 (\cite{miyamoto}, \cite{tim}),
KS/GRS 1730--312 (\cite{tsp}), GRS 1739--278 (\cite{kbor1739}) except
for a somewhat lower value of the photon index of the hard power law
component. 

No statistically significant short-term variability of the source flux
was found 
in the PCA data with an upper limit of $\la$1.5\% (2$\sigma$, for 1$^{st}$ and
2$^{nd}$ observations , $10^{-2}$--$10^2$ Hz, 2--60 keV).
This value is consistent with low values of rms of aperiodic flux
variations usually observed in the very high spectral state of black
hole binaries  (e.g. \cite{ebisawa}, \cite{miyamoto}). 
At higher energies ($\ga$15 keV), where the power law component
dominates, the upper limit on the rms of the source flux variation is 
sufficiently large -- $\sim40$\% ($2\sigma$) and therefore
inconclusive.

The spectral evolution of the source 
is illustrated by the lower panel in Fig.\ref{asmcurve} where the
hardness 
ratio derived from the ASM data is plotted. For a spectrum of the type
shown in Fig.\ref{spectrum} (the 
left panel)  the 5.0--12.2 keV to 1.3-3.0 keV  hardness ratio 
characterizes roughly relative contribution of the soft blackbody and
the power law components.  The hardness ratio increased gradually and,
within the accuracy of the ASM data, sufficiently smoothly.
Such a behavior was interrupted by a short episode of increased
hardness around TJD 10708, lasted for $\sim 5-10$ days. 
At that time the hardness ratio reached $\approx 1.0$, which is close to
the hardness ratio of the Crab spectrum, i.e. for a power law spectrum
the photon index would be $\approx 2$. At the same time the Gacaltic
Center region was observed by SIGMA telescope of GRANAT observatory
and the source was detected in hard X--rays (35--75 keV) by
GRANAT/SIGMA (\cite{sigmaiauc}, \cite{me_1755}, \cite{goldoni}). According
to the SIGMA data the power 
law approximation of the source spectrum above 40 keV gave a value of
the photon index of $\approx 3.0$. Although no spectrally resolved
data below 30 keV are available, the count rate in three ASM channels
can be used to derive a crude spectrum of the source. The ASM and
GRANAT/SIGMA data points can be roughly described by a Comptonized
cloud model with temperature kT$\sim20$ keV and optical depth $\tau \sim4$.
Simultaneously with the decrease of 
the hardness ratio observed by ASM after TJD 10708, the 35-75 keV flux
decreased as well (\cite{goldoni}, \cite{me_1755}), and according to the ASM
data the source has likely recovered the two component spectral shape with 
dominating soft component by TJD 10717 (Fig.\ref{spectrum}, the middle panel). 
The subsequent evolution of the hardness ratio is less certain
because of reduced significance of the source flux, but most likely
the ratio increased up to $\sim$1.0 by Nov. 1997.

This behavior was confirmed by the subsequent RXTE observations of XTE
J1755--324. The spectrum obtained with PCA and HEXTE during
a scan observation on Nov. 6, 1997 can be described by a power
law model with a photon index $\alpha \sim$ 2.0 at least up to $\sim$100
keV. Comparison of the two PCA spectra obtained on Nov.6 and Nov.12
shows that the decrease of the X--ray flux was accompanied by
hardening of the power law spectrum (Table \ref{param}). The spectrum
of the source obtained by PCA and HEXTE on Nov.12 is shown in
Fig.\ref{spectrum} (the right panel).

\begin{figure}
\epsfxsize=9cm 
\epsffile{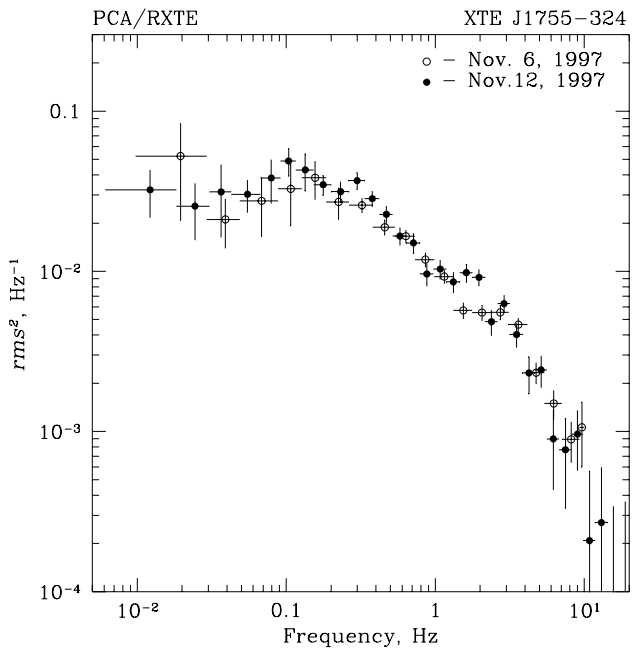} 
\caption{Power density spectra of XTE J1755--324 in two last observations.
\label{pds}}   
\end{figure}

The aperiodic variability properties of the source have changed as
well -- the values of rms in the $10^{-2}$--$10^2$ Hz frequency range
are $rms=24.04\pm1.5$\% on Nov.6 and $rms=22.4\pm1.0$\% on Nov.12,
compared to less then 1.5\% (2$\sigma$ upper limit) for Aug.1, 1997
observation. The power density spectrum has a shape typical for the
 hard spectral state of black hole binaries (Fig.\ref{pds}).

An excess is clearly seen around 6-7 keV on the PCA spectrum obtained on
Nov.12 (Fig.\ref{spectrum}). This feature can be adequately approximated by
gaussian line at the energy $6.77\pm0.16$ keV and with the width
comparable to the instrument resolution. The estimated line flux,
$\sim2.7\times10^{-4}$ phot/s/cm$^2$,  is comparable with
expected diffuse 6.7 keV line from the Galactic Ridge, mapped
by GINGA (Yamauchi \& Koyama 1993). Therefore the line flux and
equivalent width  given in the Table \ref{param} should be considered
as the upper limits on the intrinsic line parameters.

\section{Discussion}

The pattern of spectral and temporal evolutions of XTE J1755--324 bears
several features in common with other X--ray Novae, e.g. X-ray Nova
Vulpeculae 88  
(\cite{mir-kvant}), Nova Muscae 1991 (\cite{miyamoto}, \cite{tim}),
KS/GRS 1730--312 (\cite{tsp}), GRS 1739--278 (\cite{kbor1739}):
\begin{enumerate}
\item 
The shape of the X--ray light curve (short risetime,
nearly exponential decay, secondary and tertiary outbursts --
Fig.\ref{asmcurve}) is frequently observed for X--ray Novae
(\cite{tanaka}). 
\item 
The two component spectrum and the low level of aperiodic variability
of the X--ray flux observed at the peak of the light curve are well
established  features of the very high state of the black hole
candidates and are often observed at the peak of the light curve of
other X--ray Novae. 

\indent The spectrum of XTE J1755--324 
was dominated by the soft component which contributed more than 90\%
to the source luminosity. The shape of the soft component suggests
that it may originates from the optically thick part of the accretion
disk. Spectral fit by a simple multicolor blackbody disk model
indicates that  the inner radius of the optically thick disk was
sufficiently close to the compact object. However, one should bear in
mind that the value of the inner disk radius derived above depends on
the spectral model, the binary system inclination angle and assumed
source distance.

Based on the source spectrum approximated by the two-component
model consisting of a multicolor blackbody disk model and a hard power law
component, the 3--30 keV  luminosity near the peak of the light curve is
$L_{3-30 keV}\sim1.3\times10^{37}$ erg/s for 8.5 kpc distance.  
The total absorption corrected luminosity of the source in the 0.1--25.0 keV
energy band can be estimated as  $L_{0.1-25 keV}\sim1\times10^{38}$ erg/s.
This value is close to the luminosity of other black hole candidates in the
very high state, which might indicate that the assumption about the source
distance is roughly correct.

\item Anticorrelated behavior of the X--ray flux and the 5.0--12.2
keV to 1.3-3.0 keV  hardness ratio before and during some time after
the primary outburst (Fig.\ref{asmcurve}, the bottom panel). 

The hardness ratio plot shows that the
spectrum of XTE J1755--324 softened during the initial rise of the
flux. Similar behavior was observed
for other X-ray Novae -- GRS/GS 1124--64 (Nova Muscae 1991,
\cite{miyamoto}), KS/GRS 1730--312 (\cite{tsp}), GRS 1739--278
(\cite{kbor1739}), and is likely due to the increase of the relative
contribution and the absolute luminosity of the soft spectral component
caused by increase of the mass accretion rate and corresponding
increase of the disk temperature.

After the peak of the light curve  the hardness ratio increased
gradually and fairly smoothly
with a brief  excursion around TJD 10708. The gradual increase of the
hardness ratio after the primary outburst is also seen in
the spectral evolution of other
X--ray Novae. Although no spectrally resolved data is available
in the case of XTE J1755--324, we may suppose that it is related to 
decrease of the mass accretion rate and corresponding decrease of the
disk temperature (that is usually followed by a transition to the low
spectral state).

\item Transition to the low spectral state during decay.

The transition of the source to the low spectral state itself
was not traced by the pointed instruments aboard RXTE, but during
November 1997 observations the source has been found in the low
spectral state according to both spectral (Fig.\ref{spectrum}, the
right panel) and aperiodic variability (Fig.\ref{pds}) properties.  
The 4-100 keV luminosity was $L_{4-100 keV}\sim5\times10^{36}$ erg/s (Nov.
12, 1997,  PCA and HEXTE data, the HEXTE normalization was adjusted to
that of PCA).


\end{enumerate}

\medskip

A notable and apparently peculiar feature of the source evolution was
a brief excursion seen on the hardness ratio plot (Fig.\ref{asmcurve})
around TJD 10708. 
Comparison of July-Aug. 1997 and Sep. 1997 spectra of XTE J1755--324
observed with RXTE/HEXTE and GRANAT/SIGMA (Paul et. al 1997) in the
high energy domain further confirms a fact of strong spectral
evolution  (see Fig.\ref{spectrum}).  
The RXTE/HEXTE observation near the peak of the light curve on Aug. 1, 1997
gave the hard X--ray flux $F_{40-75 keV}\sim9\times10^{-11}$ ergs/s/cm$^2$.
The relative contribution of the hard power law component to the total flux
in the 3--30 keV energy band was at the level of $\sim$10\%. 
The hard X-ray flux from the source measured by GRANAT/SIGMA near the
peak of the hardness is $F_{40-75 keV}\approx5\times10^{-10}$
ergs/s/cm$^2$ (\cite{sigmaiauc}), i.e. the 40--75 keV flux from the
source increased by a factor of $\sim$5. At the same time according to
RXTE/ASM  data  the 1--12 keV flux decreased by a factor of $\sim$4--5
(Fig.\ref{asmcurve}). 
Obviously the observed spectral changes can not be accounted for in terms of
a simple scaling of the soft spectral component  according to the
luminosity change (see p.3 above). Moreover, comparison of the ASM
spectra obtained before/after and during this event
(Fig.\ref{spectrum}, the middle panel) shows clearly,
that not only luminosity in the high energy band increased but also the
luminosity below $\sim 5$ keV decreased by a factor of $\sim 2-3$. No
evidence of the soft spectral component of the type observed just
before and immediately after the event is seen in the ASM data taken
at the peak of hardness.

Assuming a power law spectrum with a photon index $\alpha=2.0$ and a high
energy cutoff (Comptonization spectrum with kT$\sim20$ keV and $\tau
\sim4$) which roughly describes ASM and SIGMA data in the 1-150 keV range,
(solid line in the middle panel in Fig.\ref{spectrum}) 
the 1.3--150 keV luminosity is $\sim2\times10^{37}$ erg/s,
which is close to typical low spectral state luminosity of Cyg X-1 and
1E1740.7--2942 (e.g. Sunyaev \& Truemper 1979, Sunyaev et. al
1991).  In general, 
the shape of the spectrum, observed with RXTE/ASM and GRANAT/SIGMA in the
middle of September 1997 is typical for the low state of the black
hole binaries except for the noticeably lower value of the Comptonization
temperature. 
A possible explanation of the event might be that it is due to 
a spectral transition to the hard state followed by a transition back
to the soft state caused by an increase of
the mass accretion rate corresponding to the tertiary maximum of the
light curve.  Such an event can ocure when the accretion rate is near
a critical level: e.g. a hard-soft-hard transition in Cyg X--1 (Zhang
et al., 1997).
Crucial data for understanding the 
nature of the event might be the timing information which unfortunately is
absent -- no pointed observations were performed by RXTE at that time. 

The pattern of the spectral and temporal evolution of XTE J1755--324 is
similar to that of ``canonical'' X--ray Novae  of dynamically proven
black hole candidates and suggests that the binary system XTE
J1755--324 also harbors a black hole.

\section{Acknowledgments}
We thank the referee, Y.Tanaka, for useful suggestions that
significantly improved the paper.
This work was partly supported by INTAS grant 93-3364-ext and RBRF grant
97-02-16264. Research has made use of data obtained through the High Energy
Astrophysics Science Archive Research Center Online Service, provided by the
NASA/Goddard Space Flight Center.

\end{document}